\documentclass[10pt,conference]{IEEEtran}

\usepackage{amsmath,amssymb,amsfonts}
\usepackage{amsthm}
\usepackage{array}
\usepackage{adjustbox}
\usepackage{balance}
\usepackage{booktabs}
\usepackage{cite}
\usepackage{color, colortbl}
\usepackage{csvsimple}
\usepackage{diagbox}
\usepackage{enumitem}
\usepackage{fancybox}
\usepackage{fontenc}
\usepackage{graphicx}
\usepackage{hyperref}
\usepackage{indentfirst}
\usepackage{lscape}
\usepackage{listings}
\usepackage{longtable}
\usepackage{makecell}
\usepackage{marvosym}
\usepackage{mathtools}
\usepackage{microtype}
\usepackage{moreverb}
\usepackage{multicol}
\usepackage{newtxmath}
\usepackage{pifont}
\usepackage{pdflscape}
\usepackage{rotating}
\usepackage{setspace}
\usepackage{soul}
\usepackage{subcaption}
\usepackage{svg}
\usepackage{threeparttable}
\usepackage{tikz}
\usepackage{tcolorbox}
\usepackage{textcomp}
\usepackage{url}
\usepackage{wasysym}
\usepackage{wrapfig}
\usepackage{xcolor}
\usepackage{xspace}
\usepackage{caption}
\usepackage{multirow}
\usepackage[capitalize,noabbrev]{cleveref}
\usepackage[ruled,vlined,lined,commentsnumbered]{algorithm2e}
\usepackage[skip=1pt,labelfont=bf]{caption}

\newcommand{\x}{\mathbf{x}}
\newcommand{\y}{\mathbf{y}}
\newcommand{\prefdata}{\mathcal D}
\let\E\relax
\newcommand{\E}[2]{\mathbb{E}_{#1}\left[{#2}\right]}
\newcommand{\piref}{\pi_\mathrm{ref}}
\makeatletter
\renewcommand{\maketag@@@}[1]{\hbox{\m@th\normalsize\normalfont#1}}%
\makeatother

\newcommand{\intuition}[1]{
\begin{tcolorbox}[colback=white,boxrule=1pt,top=0pt,bottom=0pt,left=1pt,right=2pt,top=2pt,bottom=2pt]
\em #1
\end{tcolorbox}
}

\begin{document}

\title{Learning to Align Human Code Preferences
\vspace{0.5cm}
}

\author{
\IEEEauthorblockN{Xin Yin}
\IEEEauthorblockA{
Zhejiang University \\
Hangzhou, China\\
xyin@zju.edu.cn}

\and

\IEEEauthorblockN{Chao Ni\IEEEauthorrefmark{1}}
\IEEEauthorblockA{
Zhejiang University \\
Hangzhou, China\\
chaoni@zju.edu.cn}

\and

\IEEEauthorblockN{Xiaohu Yang}
\IEEEauthorblockA{
Zhejiang University \\
Hangzhou, China\\
yangxh@zju.edu.cn}
\vspace{-1cm}
}

\maketitle

\begingroup
\renewcommand\thefootnote{\IEEEauthorrefmark{1}}
\footnotetext{
Corresponding author.
}
\endgroup

\begin{abstract}
Large Language Models (LLMs) have demonstrated remarkable potential in automating software development tasks.
While recent advances leverage Supervised Fine-Tuning (SFT) and Direct Preference Optimization (DPO) to align models with human preferences, the optimal training strategy remains unclear across diverse code preference scenarios.
This paper systematically investigates the roles of SFT and DPO in aligning LLMs with different code preferences. 
Through both theoretical analysis and empirical observation, we hypothesize that SFT excels in scenarios with objectively verifiable optimal solutions, while applying SFT followed by DPO (S\&D) enables models to explore superior solutions in scenarios without objectively verifiable optimal solutions. 
Based on the analysis and experimental evidence, we propose \textbf{A}daptive \textbf{P}reference \textbf{O}ptimization (APO), a dynamic integration approach that adaptively amplifies preferred responses, suppresses dispreferred ones, and encourages exploration of potentially superior solutions during training.
Extensive experiments across six representative code preference tasks validate our theoretical hypotheses and demonstrate that APO consistently matches or surpasses the performance of existing SFT and S\&D strategies.
Our work provides both theoretical foundations and practical guidance for selecting appropriate training strategies in different code preference alignment scenarios.
\end{abstract}


\vspace{-0.1cm}
\section{Introduction}

In recent years, Large Language Models (LLMs) have emerged as transformative tools with remarkable potential for automating diverse software development tasks. 
State-of-the-art models such as CodeLlama~\cite{roziere2023code}, WizardCoder~\cite{luowizardcoder}, and DeepSeek-Coder~\cite{deepseek-coder} have demonstrated impressive capabilities in tackling complex code generation tasks. 
These models excel at generating code snippets from natural language descriptions~\cite{peng2025soleval,du2024evaluating}, performing cross-language code translation~\cite{yin2024rectifier,pan2024lost}, automatically fixing software bugs~\cite{yin2024thinkrepair,xia2023keep}, and synthesizing comprehensive unit tests to enhance software reliability~\cite{yin2024you,wang2024hits}.
Despite these advances, a critical challenge remains in effectively aligning LLMs with human code preferences (e.g., the security and efficiency of the generated code).

To improve code generation models, a common approach is Supervised Fine-Tuning (SFT)~\cite{zhang2023instruction}, where models are trained on pairs of instructions and corresponding correct code snippets. 
Current methods generate synthetic training data through self-instruction using models like GPT-4~\cite{wang2022self,alpaca,codealpaca}.
Evol-Instruct~\cite{luowizardcoder} leverages more sophisticated prompts for enhanced data generation, while OSS-Instruct~\cite{wei2024magicoder} enables LLMs to draw inspiration from real-world code snippets to improve coding performance.
However, SFT's exclusive focus on correct examples limits its ability to teach preference discrimination, as models never encounter negative examples~\cite{hong2024orpo}.

\begin{figure}
    \centering
    \includegraphics[width=1\linewidth]{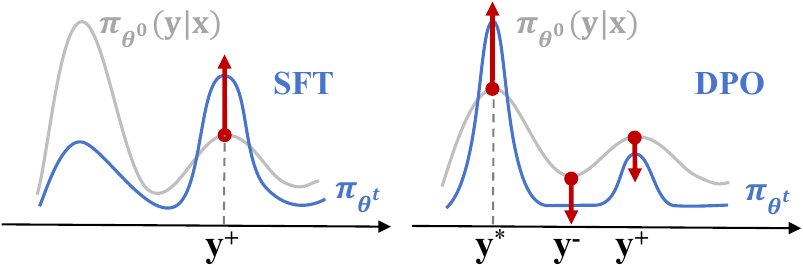}
    \caption
    {
        The updated confidence score of different algorithms
    }
    \label{fig:squeezing_effect}
    \vspace{-0.3cm}
\end{figure}

Recent research addresses these limitations through Direct Preference Optimization (DPO)~\cite{rafailov2024direct}, which aligns models using pairwise preference data. 
DPO enables models to rank outputs and select preferred solutions (e.g., more efficient code)~\cite{zhang2024codedpo,zhang2025focused,xu2024dpo}.
Prior works~\cite{pal2024smaug,rafailov2024r,tajwar2024preference,ren2024learning} reveal distinct learning behaviors between SFT and DPO, as illustrated in Fig.~\ref{fig:squeezing_effect}.
Specifically, confidence scores for both $\y^+$ and $\y^-$ gradually decrease during DPO training, while $\y^+$ confidence remains stable under SFT.
Although SFT and DPO have shown success in natural language tasks, their effectiveness across diverse code preference scenarios remains under-explored. 
Code preferences differ fundamentally from natural language, requiring objective metrics such as correctness, efficiency, and security.
Given their distinct behaviors, SFT and DPO may serve different roles depending on the specific preference scenario, complicating the selection of optimal training strategies.

In this paper, we first conduct a preliminary experiment to verify the reported phenomena~\cite{pal2024smaug,rafailov2024r,tajwar2024preference,ren2024learning} and provide a theoretical analysis to explain these behaviors.
We categorize code preference tasks into two scenarios: \ding{182} preferences with objectively verifiable optimal solutions, and \ding{183} preferences without objectively verifiable optimal solutions.
Based on these observations and theoretical analysis, we formulate the following hypotheses regarding SFT and DPO effectiveness.
In scenario \ding{182}, SFT suffices to achieve optimal solutions, while DPO provides limited additional benefits and may even harm performance.
Conversely, in scenario \ding{183}, sequential application of SFT followed by DPO (S\&D) achieves optimal performance, where SFT rapidly builds fundamental capabilities and DPO subsequently enables exploration of superior solutions.

Building on the analysis and experimental evidence, we propose \textbf{A}daptive \textbf{P}reference \textbf{O}ptimization (APO), a unified framework that dynamically integrates SFT and DPO.
APO amplifies preferred responses, suppresses dispreferred ones, and encourages exploration of superior solutions during training.

To validate these hypotheses and evaluate APO's effectiveness, we construct six code preference tasks using the APPS benchmark~\cite{hendrycks2021measuring}.
Scenario \ding{182} encompasses code correctness, security, and smell optimization, while scenario \ding{183} covers efficiency, complexity, and conciseness optimization.
Experimental results validate our hypotheses: SFT achieves superior stability in scenario \ding{182}, while (S\&D) delivers optimal performance in scenario \ding{183}.
Additionally, APO achieves comparable or superior performance across both scenarios while eliminating manual strategy selection, thereby streamlining the training pipeline.
We also compare the effectiveness and efficiency of the proposed APO framework with SFT and DPO. 
APO also maintains competitive efficiency in training time and GPU memory usage.

Our main contributions are summarized as follows:

\begin{itemize}[leftmargin=*]
    \item We theoretically analyze SFT and DPO behaviors and propose scenario-specific hypotheses for code preference alignment, providing guidance for training strategy selection.
    \item We introduce \textbf{A}daptive \textbf{P}reference \textbf{O}ptimization (APO), a unified framework that adaptively combines SFT and DPO strengths without requiring manual scenario discrimination.
    \item We empirically validate our hypotheses across six code preference tasks and demonstrate APO's superior performance while maintaining competitive training efficiency. 
\end{itemize}
\label{sec:introduction}

\section{Preliminary Study}

\begin{figure*}[htbp]
    \vspace{-0.3cm}
    \centering
    \includegraphics[width=.9\linewidth]{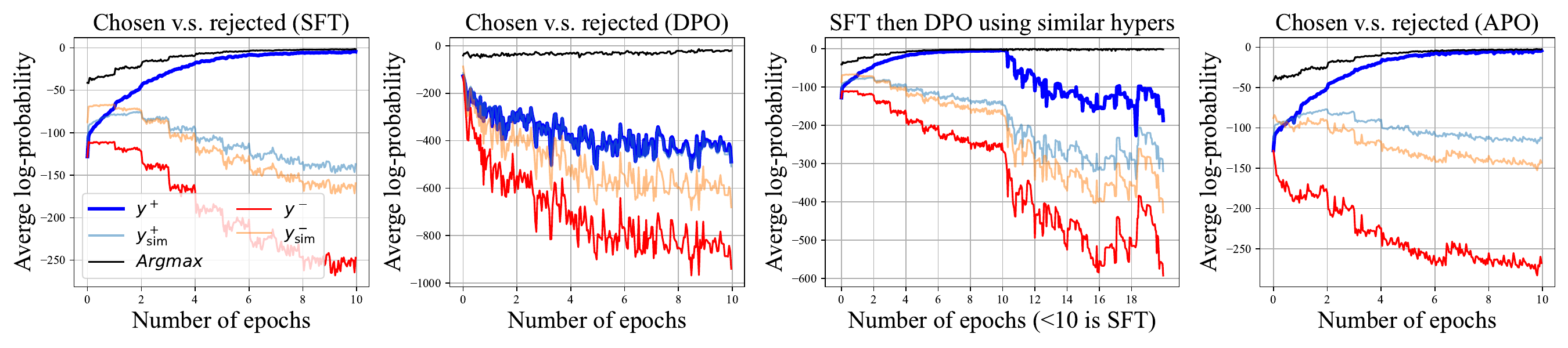}
    \caption{Learning dynamics of SFT, DPO, and APO on code correctness preference}
    \label{fig:code_correctness_preference}
\end{figure*}

\begin{figure*}[htbp]
    \centering
    \includegraphics[width=.9\linewidth]{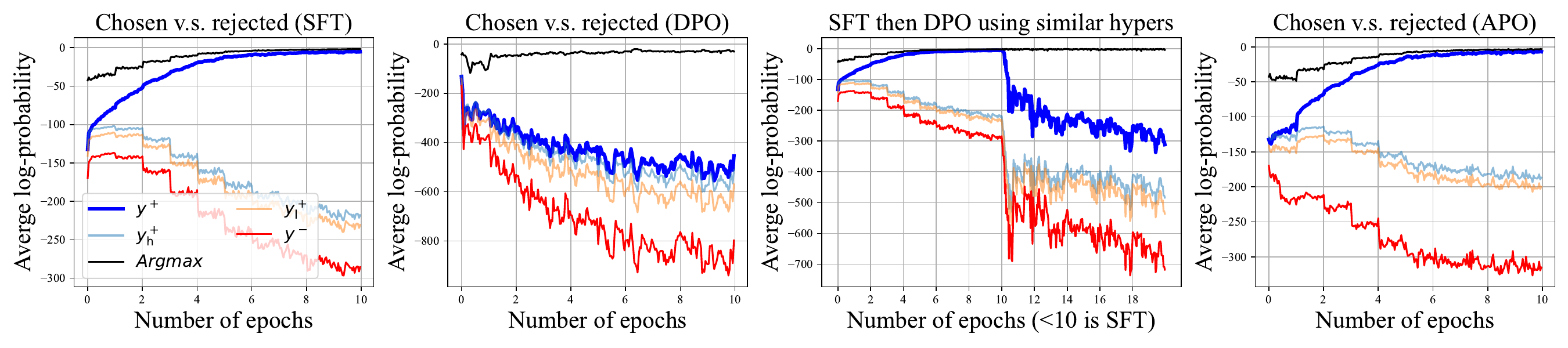}
    \caption{Learning dynamics of SFT, DPO, and APO on code efficiency preference}
    \label{fig:code_efficiency_preference}
    \vspace{-0.2cm}
\end{figure*}

\subsection{Background}
\textbf{Language Model.} 
We formalize an LLM as a conditional policy $\pi_\theta(\y \mid \x)$ parameterized by $\theta$, mapping user instructions $\x \in \mathcal{X}$ to textual responses $\y \in \mathcal{Y}$. 
Given input $\x$, the LLM generates response $\y$ auto-regressively:

\begin{equation}
\pi_\theta\left(\y\mid \x\right)=\prod_t \pi_\theta \left(y_t\mid \x, \y_{<t}\right),
\end{equation}

where $\y_t$ denotes the $t$-th token of the response, and $\y_{<t}$ are the tokens of the response before $\y_t$.

\textbf{SFT.} 
Supervised Fine-Tuning (SFT)~\cite{zhang2023instruction} trains pre-trained models to generate high-quality responses by learning from demonstration data. 
This process constructs a dataset of instruction-response pairs $\prefdata = \{(\x, \y^+)\}$ and optimizes $\pi_\theta$ via the SFT loss:

\vspace{-0.2cm}
\begin{equation}
    \mathcal{L}_\mathrm{SFT}(\pi_\theta)=-\E{(\x, \y^+)\sim\prefdata}{\log \pi_\theta(\y^+ \mid \x)}.\label{eq:sft-loss}
\end{equation}

\textbf{DPO.}
Direct Preference Optimization (DPO)~\cite{rafailov2024direct} represents a key advancement in LLM alignment, adapting models to better reflect human preferences. 
Unlike SFT, which only demonstrates desirable behaviors, DPO explicitly discourages undesirable responses through preference learning. 
DPO directly optimizes policy $\pi_\theta$ using preference data constructed as triples $\prefdata = \{(\x, \y^+, \y^-)\}$, where $\y^+$ and $\y^-$ represent preferred and dispreferred responses to prompt $\x$, respectively.
The DPO loss is defined as:

\begin{footnotesize}  
\begin{align}
    \mathcal L_\mathrm{DPO}&(\pi_\theta)=-\mathbb{E}_{(\x,\y^+,\y^-)\sim\mathcal D}\notag\\
    &\left[\log\sigma\left(\beta\left(
    \log\frac{\pi_\theta(\y^+\mid\x)}{\piref(\y^+\mid\x)}-
    \log\frac{\pi_\theta(\y^-\mid\x)}{\piref(\y^-\mid\x)}
    \right)\right)\right]\notag\\
    &=\E{(\x, \y^+, \y^-)\sim\prefdata}{\log\left(1 + \left(\frac{\piref(\y^+ \mid \x) \cdot \pi_\theta(\y^- \mid \x)}{\piref(\y^- \mid \x) \cdot \pi_\theta(\y^+ \mid \x)}\right)^{\beta}\right)}
    .
    \label{eq:dpo-loss}
\end{align}
\end{footnotesize}

\subsection{Preliminary Experiment}
Prior works~\cite{pal2024smaug,rafailov2024r,tajwar2024preference,ren2024learning} reveal distinct learning behaviors between SFT and DPO, as illustrated in Fig.~\ref{fig:squeezing_effect}.
Specifically, confidence scores for both $\y^+$ and $\y^-$ gradually decrease during DPO training, while $\y^+$ confidence remains stable under SFT.
Furthermore, they find that for some responses they track (i.e., various responses similar to $\y^+$ or $\y^-$), none of them increase during the DPO phase.
Ren et al.~\cite{ren2024learning} refer to this phenomenon observed in DPO as the \texttt{squeezing effect}, and the decreased probability mass is squeezed onto $\y^*$ (the output which was most confident before the update).

To validate their analysis in practical settings, we conduct experiments on two representative code preference tasks. 
We construct the training set $\mathcal{D}_\text{train}$ by randomly sampling 2500 examples from the datasets described in Section~\ref{sec:dataset}, covering code correctness and efficiency preferences.
Each example comprises three components: a prompt $\x$, a preferred response $\y^+$, and a dispreferred response $\y^-$. 
During training, SFT uses only $\x$ and $\y^+$, while DPO leverages all three components. 
We repeat the experiments on two series of models: \texttt{DeepSeek-Coder 1.3B}~\cite{deepseek-coder} and \texttt{Qwen2.5-Coder 0.5B/1.5B}~\cite{hui2024qwen2}.

To analyze learning dynamics in detail, we construct a probing dataset $\mathcal{D}_\text{prob}$ by sampling 500 examples from $\mathcal{D}_\text{train}$ and augmenting each with additional response variants. 
For code correctness task, we include responses similar to the original preferred and dispreferred responses, denoted as $\y^+_{\text{sim}}$ and $\y^-_{\text{sim}}$. 
For code efficiency task, we add responses with higher efficiency ($\y^+_h$) and lower efficiency ($\y^+_l$) than $\y^+$, reflecting real-world scenarios where the preferred response may not represent the optimal solution.
We fine-tune the models for several epochs, evaluating predictions on all responses in $\mathcal{D}_\text{prob}$ every 100 updates.
For each response type, we compute the average log-probability across all 500 examples as a measure of model confidence.
This allows us to track how $\log\pi_{\theta^t}(\y \mid \x)$ evolves for different response categories throughout training.

\textbf{Learning dynamics of SFT and DPO.}
As demonstrated in the first panel of Fig.~\ref{fig:code_correctness_preference} and Fig.~\ref{fig:code_efficiency_preference}, SFT consistently increases the model's confidence in $\y^+$ throughout training, which is expected given that the optimization directly applies ``pull-up'' pressure to the preferred response. 
This increase in $\y^+$ probability naturally diminishes confidence in all other responses $\y\neq\y^+$, as the model's predicted probabilities across the entire $\mathcal{Y}$-space must normalize to one.
In contrast, the second panel reveals that DPO exhibits markedly different behavior: confidence scores for all $\y$ responses decrease during training, with $\y^-$ and its similar variants experiencing the most significant drops, as the decreased probability mass is squeezed onto $\y^*$ (Argmax).
Notably, the third panel demonstrates that even when DPO is applied after SFT training, the confidence in $\y^+$ still decreases.
These empirical observations align consistently with the phenomena reported by Ren et al.~\cite{ren2024learning}.
\label{sec:preliminary}

\section{Insights of SFT and DPO in Practical Usage}

\crefname{equation}{Eq.}{Eqs.}
\crefname{theorem}{Theorem}{Theorems}
\Crefname{theorem}{Theorem}{Theorems}
\theoremstyle{plain}
\newtheorem{theorem}{Theorem}
\newtheorem{proposition}[theorem]{Proposition}
\newtheorem{lemma}[theorem]{Lemma}
\newtheorem{corollary}[theorem]{Corollary}
\theoremstyle{definition}
\newtheorem{definition}[theorem]{Definition}
\newtheorem{assumption}[theorem]{Assumption}
\theoremstyle{remark}
\newtheorem{remark}[theorem]{Remark}

This section highlights the respective strengths of DPO and SFT in aligning with human code preferences.
We first conduct a theoretical analysis of how $\y^+$ and $\y^-$ probabilities evolve when the training objectives of SFT and DPO are optimized.
We then present a synthetic example to further illustrate their differences.
Building upon the preliminary experiments from Section~\ref{sec:preliminary}, we systematically evaluate the effectiveness of both methods across different scenarios of human code preferences.
Finally, based on the analysis and experimental evidence, we design a new training method to align with various human code preferences.

\subsection{Theoretical Analysis}
\begin{theorem}
\label{thm:main}
We analyze the objectives of SFT and DPO under ideal optimization conditions to understand their fundamental differences.
Assume that both optimization processes converge to their respective global minima.

Given a preference dataset $\mathcal{D}$, let $\Pi_\mathrm{SFT}$ and $\Pi_\mathrm{DPO}$ denote the sets of optimal policies obtained by minimizing the SFT loss in Eq.~\ref{eq:sft-loss} and the DPO loss in Eq.~\ref{eq:dpo-loss}, respectively.
SFT directly maximizes the likelihood of preferred responses, while DPO optimizes relative preferences between response pairs, enhancing the model's ability to distinguish superior responses from multiple candidates.
Under perfect optimization, we establish the following theoretical results:
\ding{182} When $\mathcal{L}_\mathrm{SFT}(\pi_\theta) \to 0$, we have $\pi_\theta(\y^+ \mid \x) \to 1$ and consequently $\pi_\theta(\y^- \mid \x) \to 0$;
\ding{183} When $\mathcal{L}_\mathrm{DPO}(\pi_\theta) \to 0$, we have $\pi_\theta(\y^- \mid \x) \to 0$, but $\pi_\theta(\y^+ \mid \x)$ does not necessarily approach 1.
\end{theorem}

\noindent \textbf{\textit{Proof.}}
We first prove the conclusion \ding{182}. 
The SFT objective minimizes the negative log-likelihood of preferred responses, thereby encouraging the model to assign maximal probability to these responses.  
Formally,

\vspace{-0.2cm}
\begin{align}
    \lim_{\x \to \x_0} \mathcal{L}_\mathrm{SFT}(\pi_\theta)&=\lim_{\x \to \x_0} -\E{(\x, \y^+)\sim\prefdata}{\log \pi_\theta(\y^+ \mid \x)} = 0\notag\\
    &\Rightarrow \pi_\theta(\y^+ \mid \x) \to 1\notag\\
    &\Rightarrow \pi_\theta(\y^- \mid \x) \to 0.
\end{align}

This result follows directly from the properties of the cross-entropy loss: the loss approaches zero if and only if the model assigns a probability approaching one to the target label (i.e., the preferred response $\y^+$).

\begin{table}[htbp!]
    \centering
    \caption{A counter-example with three actions}
    \label{tab:dpo-counterexample}
    \begin{tabular}{c|ccc}
    \toprule
        Policy & $\y_1$ & $\y_2$ & $\y_3$\\
        \midrule
        $\piref$ & $0.3$ & $0.2$ & $0.5$ \\
        $\mathcal{D}_\mathrm{pref}$ & \multicolumn{3}{c}{$\{(\y^+=\y_1,\y^-=\y_2)\}$} \\
        \midrule
        $\pi_\mathrm{DPO}, \mathcal{L}_\mathrm{DPO} = 0 $ & $0.2$ & $0.0$ & $0.8$ \\
        $ \quad \quad \mathcal{L}_\mathrm{SFT} \neq 0 $ & 0.2 & 0.0 & 0.8 \\
        \midrule
        $\pi_\mathrm{SFT}, \mathcal{L}_\mathrm{SFT} = 0 $ & 1.0 & 0.0 & 0.0 \\
        \bottomrule
    \end{tabular}
\end{table}

Next, we present a counter-example in Table~\ref{tab:dpo-counterexample} to demonstrate the conclusion \ding{183}.
Consider a scenario with three possible responses $\{\y_1,\y_2,\y_3\}$, where our preference dataset $\mathcal{D}$ contains only a single comparison pair $(\y^+,\y^-)=(\y_1,\y_2)$. 
Let $a = \piref(\y^+ \mid \x)$ and $b = \piref(\y^- \mid \x)$ denote the reference policy probabilities. 
From the DPO loss in Eq.~\ref{eq:dpo-loss}, when the loss approaches zero, we have:

\vspace{-0.3cm}
\begin{footnotesize}
\begin{align}
    \lim_{\x \to \x_0}
    \mathcal{L}_\mathrm{DPO}(\pi_\theta) &= \lim_{\x \to \x_0}
    \mathbb{E}_{(\x, \y^+, \y^-) \sim \mathcal{D}} \left[\log \left( 1 + \left( \frac{b \cdot \pi_\theta(\y^- \mid \x)}{a \cdot \pi_\theta(\y^+ \mid \x)} \right)^\beta \right)\right]=0\notag\\
    &\Rightarrow \frac{b \cdot \pi_\theta(\y^- \mid \x)}{a \cdot \pi_\theta(\y^+ \mid \x)} \to 0.
\end{align}
\end{footnotesize}
\vspace{0.1cm}

Since $a$ and $b$ are fixed positive constants for all $(\x, \y^+, \y^-) \in \mathcal{D}$, the above condition directly implies $\pi_\theta(\y^- \mid \x) \to 0$. 
However, minimizing $\mathcal{L}_\mathrm{DPO}(\pi_\theta)$ to zero does not require $\pi_\theta(\y^+ \mid \x) \to 1$. 
To illustrate this key difference, consider the optimal DPO policy shown in the third row of \cref{tab:dpo-counterexample}, which assigns probability 0.2 to $\y_1$ and 0.8 to $\y_3$.
This policy achieves $\mathcal{L}_\mathrm{DPO} = 0$ but would be impossible under SFT, which enforces $\pi_\mathrm{SFT}(\y_1 \mid \x) = 1$ according to Eq.~\ref{eq:sft-loss}.

\subsection{Analysis in Human Code Preferences}
In this section, we analyze the objectives of SFT and DPO in practical code generation scenarios. 
We categorize the human code preferences into two cases: 

\begin{itemize}[leftmargin=*]
\item \textbf{\ding{182} Preferences with objectively verifiable optimal solutions.} 
These correspond to tasks where a definitive correct or optimal solution can be determined through objective evaluation. 
Code correctness exemplifies this category, as solutions can be rigorously verified against specifications or comprehensive test suites to confirm functional correctness.
\item \textbf{\ding{183} Preferences without objectively verifiable optimal solutions.} 
These involve tasks where optimality is subjective or difficult to ascertain. 
Code efficiency represents this category, as determining the most efficient solution is often challenging—superior alternatives may exist but remain undiscovered or require prohibitive costs to identify.
\end{itemize}

For scenario \ding{182}, we demonstrate that SFT is more effective than DPO. 
In such cases, the model can be optimized directly using SFT alone, without requiring additional DPO training, which may provide limited additional benefits and may even harm performance. 
Consider the example in \autoref{tab:dpo-counterexample}, where the optimization objective is to maximize code correctness (i.e., pass rate). 
Any solution that successfully passes all test cases represents an objectively optimal solution.
As illustrated in the second and third panels of Fig.~\ref{fig:code_correctness_preference}, DPO training reduces the probabilities assigned to both $\y^+$ and $\y^-$, squeezing the probability mass onto $y^*$. 
The third row of \autoref{tab:dpo-counterexample} demonstrates a DPO-optimal policy that assigns a probability of 0.2 to $\y_1$ and 0.8 to $\y_3$. 
Under this configuration, $\pi_\mathrm{DPO}$ fails to achieve superior performance and actually diminishes the likelihood of generating $\y_1$, while $\pi_\mathrm{SFT}$ guarantees a probability of 1.

For scenario \ding{183}, the preferred response $\y^+$ in the preference dataset may not represent the optimal solution (e.g., a more efficient response $\y$ may exist but be undiscovered by humans). 
In this context, SFT and DPO provide complementary advantages: SFT rapidly elevates the model's baseline capabilities to match the quality of demonstrated solutions in the preference dataset, while DPO enables the model to potentially surpass these demonstrated solutions through continued exploration. 
As illustrated in the second and third panels of Fig.~\ref{fig:code_efficiency_preference}, the probability mass is still squeezed onto $\y^*$. 
When $\y^*$ represents a more efficient solution, DPO can facilitate the model's achievement of a higher performance ceiling.

\intuition{{\bf Hypothese.}
(1) In scenario \ding{182}, SFT suffices to achieve optimal solutions, while DPO provides limited additional benefits and may even harm performance.
(2) In scenario \ding{183}, sequential application of SFT followed by DPO (S\&D) achieves optimal performance, where SFT rapidly builds fundamental capabilities and DPO subsequently enables exploration of superior solutions.
}

Based on the above analysis, we formulate these hypotheses and will empirically validate them in Section~\ref{sec:results}. 
Our goal is to investigate whether our hypotheses hold in these two scenarios and to guide training strategy selection that enables the model to achieve superior performance.

\subsection{Adaptive Preference Optimization}
While SFT effectively increases the probability assigned to preferred responses ($\y^+$), DPO enables exploration of superior solutions and effectively reduces the probability of dispreferred responses ($\y^-$), though it may also inadvertently decrease the probability of preferred responses ($\y^+$).
Since SFT and DPO demonstrate distinct advantages across different scenarios, we propose a unified approach that eliminates the need for manual scenario discrimination while simultaneously boosting the probability of preferred responses ($\y^+$), suppressing dispreferred responses ($\y^-$), and encouraging exploration of potentially superior solutions during training.
We term this dynamic integration framework \textbf{A}daptive Preference Optimization (APO), which adaptively combines both methods through a dynamic loss:

\vspace{-0.2cm}
\begin{equation}
    \mathcal{L}_{\mathrm{APO}}(\pi_\theta)
    = \alpha \,\mathcal{L}_{\mathrm{DPO}}(\pi_\theta)\;+\;(1 - \alpha)\,\mathcal{L}_{\mathrm{SFT}}(\pi_\theta),
\end{equation}

where $\alpha = e^{\frac{1}{T} \sum_{t=1}^{T} \log \pi_\theta(\y_t \mid \y_{<t})}$ denotes the geometric mean of the model's probability of generating the sequence.
When $\alpha$ is low, the SFT term dominates, enabling rapid convergence toward $\y^+$.
As $\alpha$ increases, the DPO term takes precedence, guiding the model to refine its policy and exploit more nuanced preference information.
In the early stages of training, the model rarely assigns high probability to any preferred response, resulting in \(\mathcal{L}_{\mathrm{APO}}\approx \mathcal{L}_{\mathrm{SFT}}\). 
As learning progresses and positive examples gain probability mass, \(\mathcal{L}_{\mathrm{APO}}\approx \mathcal{L}_{\mathrm{DPO}}\), enabling the model to continue exploring superior solutions.
This adaptive strategy effectively combines the advantages of both SFT and DPO. 
It leverages the direct maximum-likelihood signal to rapidly initiate learning, and as the model's outputs increasingly align with human preferences, it further reduces the probability of dispreferred responses.
As demonstrated in Fig.~\ref{fig:code_correctness_preference} and Fig.~\ref{fig:code_efficiency_preference}, APO effectively increases the probability of generating $\y^+$ in both scenarios while rapidly reducing the probability of $\y^-$ during the early training stages. 
In the code efficiency preference scenario, APO achieves a higher probability for $\y^+_h$ compared to SFT alone, demonstrating that the model continues to explore superior solutions.
\label{sec:understand}

\section{Experimental Design}

In this section, we conduct comprehensive experiments on preference datasets to investigate the effectiveness of SFT and DPO across diverse code preference scenarios. 
We present our studied code preference scenarios, constructed datasets, studied LLMs, evaluation metrics, and experimental settings.

\subsection{Studied Code Preference Scenarios}
\label{lab:preference}

For preferences with objectively verifiable optimal solutions, we consider three types of human code preference scenarios:
\begin{itemize}[leftmargin=*]
\item \textbf{Code Correctness.} 
Ensuring that code executes correctly and produces expected outputs according to specified input-output requirements.
\item \textbf{Code Security.} 
Beyond correctness, humans prefer generated code with minimal vulnerabilities, ensuring safe handling of sensitive data and protection against security threats.
\item \textbf{Code Smell.}
It refers to any symptoms in the code that may lead to deeper issues. 
Humans prefer well-structured code that establishes best practices and maintains high readability.
\end{itemize}

For preferences without objectively verifiable optimal solutions, we examine three additional code preference scenarios:
\begin{itemize}[leftmargin=*]
\item 
\textbf{Code Efficiency.}
When multiple solutions achieve identical functionality, humans favor more efficient solutions that execute faster and consume fewer computational resources.
\item 
\textbf{Code Complexity.}
This evaluates the complexity of the code by measuring the number of linearly independent paths through the program.
Humans prefer simpler, more understandable code that facilitates maintenance and testing.
\item 
\textbf{Code Conciseness.} 
Humans prefer concise codes for their enhanced readability and comprehensibility. 
Concise code typically reduces maintenance overhead, minimizes error introduction, and promotes effective developer collaboration.
\end{itemize}

\subsection{Constructed Dataset}
\label{sec:dataset}
We construct our preference dataset using the widely adopted APPS dataset~\cite{hendrycks2021measuring}. 
The APPS dataset comprises 10,000 coding problems sourced from various open-access coding platforms, including Codeforces and Kattis. 
These problems span difficulty levels from introductory to collegiate competition level, designed to assess both coding proficiency and problem-solving capabilities.
Each problem is presented in unrestricted natural language, closely reflecting real-world programming scenarios. 
The dataset includes 131,836 test cases for solution verification and 232,444 ground-truth solutions authored by human programmers.
On average, each problem contains 293.2 words and is accompanied by comprehensive test cases (21.2 per problem in the test set), specifically designed to rigorously evaluate program functionality. 
The dataset is evenly partitioned into training and test sets, each containing 5,000 problems.

Since problems in the APPS training set contain relatively few test cases, making functional correctness verification challenging, we perform three systematic steps on the test set to construct high-quality preference datasets. 
The statistics for each step are presented in Table~\ref{tab:dataset}.

\textbf{Step 1: }We randomly sample 10\% of problems from the test set to create a new test set, resulting in 500 test problems.

\textbf{Step 2: }For the remaining 4,500 problems, we utilize both the original solutions from the APPS dataset and generate additional high-quality solutions using three state-of-the-art code generation models: \texttt{DeepSeek-Coder 33B}~\cite{deepseek-coder}, \texttt{CodeLlama 34B}~\cite{roziere2023code}, and \texttt{Qwen2.5-Coder 32B}~\cite{hui2024qwen2}. 
We generate 10 solutions per problem from each model, ensuring diverse solution coverage.

\textbf{Step 3: }All generated solutions undergo rigorous evaluation using the corresponding test cases from the APPS dataset.
Solutions failing to pass all test cases are systematically filtered out, yielding a total of 120,833 functionally correct solutions.

\begin{table}[htbp]
  \centering
  \caption{Statistics of the constructed dataset}
  \resizebox{.8\linewidth}{!}
  {
    \begin{tabular}{lccc}
    \toprule
    \textbf{Dataset} & \textbf{\# Problem} & \textbf{\makecell{\# Solution\\(Preference)}} & \textbf{\makecell{Optimal\\Solution}} \\
    \midrule
    Test Set & 500 & - & - \\
    Augmented & 4,500 & 120,833 & - \\
    \midrule
    Correctness & 3,868  & 3,868 & \ding{51} \\
    Security & 3,145 & 3,145 & \ding{51} \\
    Smell & 2,809  & 9,077 & \ding{51} \\
    Efficiency & 3,358 & 3,358 & \ding{55} \\
    Complexity & 3,392 & 3,392 & \ding{55} \\
    Conciseness & 3,349 & 3,349 & \ding{55} \\
    \bottomrule
    \end{tabular}%
  }
  \label{tab:dataset}%
\end{table}%

To evaluate the effectiveness of SFT, DPO, and APO, we construct specialized preference training sets for each code preference scenario:

\begin{itemize}[leftmargin=*]
  \item 
  \textbf{Code correctness:} We employ a PageRank-inspired~\cite{lawrence1998pagerank} iterative algorithm to rank code solutions. 
  Initially, each solution receives a validation score of 1.
  For each coding problem, we compute scores for solutions and test cases based on their mutual performance: test cases with fewer passing solutions receive higher scores, while those with more passing solutions receive lower scores. 
  Conversely, solutions passing more test cases receive higher scores than those passing fewer test cases.
  These scores are iteratively updated over T = 5 iterations based on validation performance.
  Let $P \in \{0,1\}^{n \times m}$ denote the binary pass matrix where $P_{ij} = 1$ indicates solution $i$ passes test case $j$, and $F = \mathbf{1} - P$ represents the corresponding fail matrix. 
  Let $\mathbf{s} \in \mathbb{R}^n$ and $\mathbf{t} \in \mathbb{R}^m$ denote the score vectors for solutions and test cases, respectively. The iterative update process follows:

  \vspace{-0.3cm}
  \begin{align}
  \label{eq:bidirectional-ranking}
  \mathbf{s}^{(k+1)} &= \alpha \times P \times \mathbf{t}^{(k)} + (1 - \alpha) \times \mathbf{s}^{(k)} \notag \\
  \mathbf{t}^{(k+1)} &= \alpha \times F^\top \times \mathbf{s}^{(k)} + (1 - \alpha) \times \mathbf{t}^{(k)},
  \end{align}
  \vspace{0.01cm}
  
  \vspace{-0.2cm}
  where $\alpha \in (0,1)$ is a damping factor controlling the balance between newly propagated scores and previous scores. 
  After each iteration, $\mathbf{s}^{(k+1)}$ and $\mathbf{t}^{(k+1)}$ are normalized to ensure $\sum_i s_i = n$ and $\sum_j t_j = m$.

  For each problem, we select the highest and lowest scoring solutions from all successfully compiled codes, yielding 3,868 preference pairs.
  
  \item \textbf{Code security:} Since APPS primarily focuses on algorithmic problems and rarely contains inherent security vulnerabilities, we utilize injection methods from ProsSec~\cite{xu2024prosec} to systematically introduce vulnerabilities into solutions. 
  We construct preference pairs by contrasting vulnerable and non-vulnerable solution variants, resulting in 3,145 preference pairs. 
  Evaluation is conducted using the CyberSec~\cite{wan2024cyberseceval} benchmark.
  
  \item \textbf{Code smell:} Following previous works~\cite{van2021prevalence,siddiq2022empirical}, we employ the Pylint static analysis tool for smell detection. 
  We apply Pylint to each solution, identifying code smells requiring refactoring.
  For each detected smell type, we randomly pair the corresponding solution with a smell-free counterpart, generating 9,077 preference pairs.
  
  \item \textbf{Code efficiency:} Following the methodology of~\cite{peng2025coffe}, we utilize the Cirron~\cite{cirron} library to measure CPU instruction counts. 
  Unlike execution time measurements, which can be influenced by external factors, CPU instruction counts provide stable, deterministic performance metrics.
  We select the most and least efficient solutions for each problem, creating 3,358 preference pairs.
  
  \item \textbf{Code complexity:} We employ the Radon~\cite{radon} tool to calculate cyclomatic complexity, selecting solution pairs with maximum and minimum complexity scores. 
  This process yields 3,392 preference pairs.

  \item \textbf{Code conciseness:} We calculate token lengths for each solution and pair the longest and shortest implementations per problem, resulting in 3,349 preference pairs.
  \end{itemize}

\subsection{Studied Large Language Models}
We evaluate four series of open-source LLMs: 
\texttt{Llama 3.2 1B}~\cite{meta2024llama}, \texttt{DeepSeek-Coder 1.3B/6.7B}~\cite{deepseek-coder}, \texttt{Magicoder 6.7B}~\cite{wei2024magicoder}, and \texttt{Qwen2.5-Coder 1.5B/7B}~\cite{hui2024qwen2}. 
Parameter count constraints are imposed by our available computational resources (4 × NVIDIA A800 GPUs).

\subsection{Evaluation Metrics}

Inspired by \textit{Pass@k}~\cite{humaneval} and \textit{Efficient@k}~\cite{peng2025coffe}, we further expend \textit{\{Preference\}@k} to comprehensively evaluate various code preference scenarios.
The definition of \textit{\{Preference\}@k} is formalized in Eq.~\ref{eq:eff}.

\begin{equation}\label{eq:eff}
  \text{\{Preference\}@k}:=\underset{\text { Problems }}{\mathbb{E}}\left[1-\frac{\binom{n-c_p}{k}}{\binom{n}{k}}\right],
\end{equation}

where $c_p$ represents the number of solutions that pass all test cases while satisfying the specific preference criterion: being free of code smells, more efficient, less complex, or more concise than the APPS-provided baseline solution.
With values ranging from 0 to \textit{Pass@k}, \textit{\{Preference\}@k} integrates functional correctness with task-specific evaluation for comprehensive code quality assessment.
For specific criteria, we define corresponding metrics: \textit{Clean@k} for code smell, \textit{Simple@k} for code complexity, and \textit{Concise@k} for code conciseness.

Beyond \textit{\{Preference\}@k}, we evaluate \textit{\{Preference\} Rate}, defined as the proportion of solutions that pass all test cases while satisfying the specific preference criterion.
We define corresponding metrics: \textit{Clean Rate}, \textit{Efficiency Rate}, \textit{Simplicity Rate}, and \textit{Conciseness Rate} for their respective criteria.
For code security, we report \textit{Security Rate} as measured on the CyberSec~\cite{wan2024cyberseceval} benchmark.

\subsection{Implementation}
\label{sec:implementation}
We implement the generation pipeline in Python using PyTorch~\cite{pytorch} implementations of \texttt{Llama 3.2}, \texttt{DeepSeek-Coder}, \texttt{Magicoder}, and \texttt{Qwen2.5-Coder}. 
Model weights are loaded and outputs generated via Hugging Face~\cite{huggingface} libraries.
For SFT, DPO, and APO training, we select models with at most 7B parameters due to computational constraints. 
To prevent overfitting, we limit training to 5 epochs. 
For evaluation, both original and fine-tuned models generate 5 solutions per test problem. 
Experiments are conducted on a 32-core workstation equipped with Intel(R) Xeon(R) Platinum 8358P CPU @ 2.60GHz, 2TB RAM, and 4×NVIDIA A800 80GB GPUs, running Ubuntu 20.04.6 LTS.
\label{sec:experiment}

\section{Results}

To investigate the effectiveness of SFT, DPO, and APO in aligning human code preferences, we conduct comprehensive experiments addressing the following three research questions:

\begin{itemize}[leftmargin=*]
\item \textit{RQ-1 How well does SFT and DPO perform on scenarios with objectively verifiable optimal solutions?}
\item \textit{RQ-2 How well does SFT and DPO perform on scenarios without objectively verifiable optimal solutions?}
\item \textit{RQ-3 Can APO achieve comparable performance to SFT and S\&D on code preference scenarios?}
\end{itemize}

\subsection{RQ-1: Comparable Study of SFT and DPO on Scenario \ding{182}}

\noindent
\textbf{Objective.}
With the continuous advancement of deep learning technologies, LLMs have become essential tools for code generation. 
SFT and DPO are two prominent training strategies that have demonstrated promising results in various code generation tasks~\cite{zhang2023instruction,zhang2024codedpo,yin2024multitask}. 
In Section~\ref{sec:understand}, we hypothesize that for scenarios \ding{182}, SFT is sufficient to achieve optimal performance, while DPO provides limited additional benefits and may even harm performance
This research question aims to provide a comprehensive empirical evaluation of SFT and DPO across different code preference tasks (i.e., code correctness, code security, and code smell).
Through this empirical evaluation, we seek to validate our hypothesis and determine which strategy offers superior practical advantages.

\noindent
\textbf{Experimental Design.}
We establish baselines using various original LLMs, including \texttt{Llama 3.2 1B} (referred to as Llama), \texttt{DeepSeek-Coder 1.3B/6.7B} (referred to as DeepSeek), \texttt{Magicoder 6.7B} (referred to as Magicoder), and \texttt{Qwen2.5-Coder 1.5B/7B} (referred to as Qwen). 
These models are subsequently fine-tuned using both SFT and DPO strategies, leveraging the datasets constructed in Section~\ref{sec:dataset}.
Additionally, we employ a sequential training strategy, performing SFT followed by DPO (referred to as S\&D).
Each example in $\mathcal{D}$ consists of three components: the prompt (or question) $\x$, the preferred response $\y^+$, and the less preferred response $\y^-$. 
SFT is fine-tuned using $\x$ and $\y^+$, while DPO utilizes all three components.
For code correctness, we report the overall \textit{Pass@5} metric and \textit{Pass@5} metrics segmented by difficulty levels: competition, interview, and introductory. 
For code smell, we calculate \textit{Clean@5} and \textit{Clean Rate}. 
For code security, we report the \textit{Security Rate} as measured on the CyberSec~\cite{wan2024cyberseceval} benchmark.

\noindent
\textbf{Results.}
We present and analyze the results for code correctness, security, and smell preferences, respectively.

\underline{\textbf{Code Correctness Preference.}}
The evaluation results for code correctness are presented in Table~\ref{tab:correctness}, with the best performances highlighted in bold.
Based on these results, we observe that the SFT training strategy outperforms the DPO training strategy, achieving superior results across most metrics.
Specifically, SFT achieves Pass@5 scores ranging from 6.2\% to 27.0\%, representing improvements of 1.0\% to 6.2\% over the original models. 
In contrast, DPO yields a lower Pass@5 range of 2.6\% to 25.4\%.
Notably, for DeepSeek-Coder 1.3B, DPO not only fails to enhance the original model's capabilities but actually reduces its functional accuracy.
Furthermore, applying DPO after an initial SFT phase (S\&D) does not consistently yield significant advantages over SFT alone; in some cases, it can even diminish SFT's effectiveness. 
For instance, on the Qwen2.5-Coder 7B model, the S\&D strategy achieves a Pass@5 of only 25.8\%, which is inferior to the 27.0\% achieved by SFT alone.

\begin{table}[htbp]
\centering
\caption{RQ-1: The effectiveness of SFT and DPO in improving code correctness}
\label{tab:correctness}
\resizebox{.85\linewidth}{!}
{
    \begin{tabular}{lc|rrrr}
    \toprule
    \textbf{LLM} & \textbf{Method} & \textbf{Pass@5} & \textbf{Comp.} & \textbf{Inter.} & \textbf{Intro.} \\
    \midrule
    \multirow{3.5}[2]{*}{DeepSeek 1.3B} & Ori. & 8.6\% & 1.0\% & 6.6\% & \cellcolor[rgb]{ .886,  .937,  .855}\textbf{22.0\%} \\
    & SFT & \cellcolor[rgb]{ .886,  .937,  .855}\textbf{9.6\%} & 0.0\% & \cellcolor[rgb]{ .886,  .937,  .855}\textbf{8.6\%} & \cellcolor[rgb]{ .886,  .937,  .855}\textbf{22.0\%} \\
    & DPO & 8.0\% & \cellcolor[rgb]{ .886,  .937,  .855}\textbf{3.1\%} & 6.3\% & 18.0\% \\
    & S\&D & 9.0\% & 2.0\% & 7.9\% & 19.0\% \\
    \midrule
    \multirow{3.5}[2]{*}{Qwen 1.5B} & Ori. & 9.4\% & 1.0\% & 6.6\% & 26.0\% \\
    & SFT & \cellcolor[rgb]{ .886,  .937,  .855}\textbf{14.8\%} & \cellcolor[rgb]{ .886,  .937,  .855}\textbf{3.1\%} & \cellcolor[rgb]{ .886,  .937,  .855}\textbf{13.2\%} & \cellcolor[rgb]{ .886,  .937,  .855}\textbf{31.0\%} \\
    & DPO & 11.0\% & 0.0\% & 9.6\% & 26.0\% \\
    & S\&D & {14.6\%} & \cellcolor[rgb]{ .886,  .937,  .855}\textbf{3.1\%} & {12.9\%} & \cellcolor[rgb]{ .886,  .937,  .855}\textbf{31.0\%} \\
    \midrule
    \multirow{3.5}[2]{*}{Llama 1B} & Ori. & 1.4\% & 0.0\% & 1.3\% & 3.0\% \\
    & SFT & \cellcolor[rgb]{ .886,  .937,  .855}\textbf{6.2\%} & \cellcolor[rgb]{ .886,  .937,  .855}\textbf{1.0\%} & \cellcolor[rgb]{ .886,  .937,  .855}\textbf{5.3\%} & \cellcolor[rgb]{ .886,  .937,  .855}\textbf{14.0\%} \\
    & DPO & 2.6\% & 0.0\% & 1.7\% & 8.0\% \\
    & S\&D & 3.8\% & 0.0\% & 3.0\% & 10.0\% \\
    \midrule
    \multirow{3.5}[2]{*}{DeepSeek 6.7B} & Ori. & 17.8\% & \cellcolor[rgb]{ .886,  .937,  .855}\textbf{5.1\%} & 15.2\% & 38.0\% \\
    & SFT & \cellcolor[rgb]{ .886,  .937,  .855}\textbf{22.0\%} & 4.1\% & 18.5\% & \cellcolor[rgb]{ .886,  .937,  .855}\textbf{50.0\%} \\
    & DPO & 18.6\% & \cellcolor[rgb]{ .886,  .937,  .855}\textbf{5.1\%} & 15.6\% & 41.0\% \\
    & S\&D & \cellcolor[rgb]{ .886,  .937,  .855}\textbf{22.0\%} & {4.1\%} & \cellcolor[rgb]{ .886,  .937,  .855}\textbf{20.5\%} & 44.0\% \\
    \midrule
    \multirow{3.5}[2]{*}{Qwen 7B} & Ori. & 20.8\% & 4.1\% & 19.9\% & 40.0\% \\
    & SFT & \cellcolor[rgb]{ .886,  .937,  .855}\textbf{27.0\%} & \cellcolor[rgb]{ .886,  .937,  .855}\textbf{13.3\%} & \cellcolor[rgb]{ .886,  .937,  .855}\textbf{25.5\%} & \cellcolor[rgb]{ .886,  .937,  .855}\textbf{45.0\%} \\
    & DPO & 25.4\% & 12.2\% & 24.2\% & 42.0\% \\
    & S\&D & 25.8\% & 10.2\% & 24.5\% & \textbf{45.0\%} \\
    \midrule
    \multirow{3.5}[2]{*}{Magicoder 6.7B} & Ori. & 16.0\% & 5.1\% & 13.9\% & 33.0\% \\
    & SFT & \cellcolor[rgb]{ .886,  .937,  .855}\textbf{22.0\%} & 5.1\% & \cellcolor[rgb]{ .886,  .937,  .855}\textbf{19.5\%} & \cellcolor[rgb]{ .886,  .937,  .855}\textbf{46.0\%} \\
    & DPO & 18.2\% & 4.1\% & 15.6\% & 40.0\% \\
    & S\&D & 20.6\% & \cellcolor[rgb]{ .886,  .937,  .855}\textbf{7.1\%} & 17.5\% & 43.0\% \\
    \bottomrule
    \end{tabular}
    }
\end{table}

\underline{\textbf{Code Security Preference.}}
The effectiveness of SFT and DPO in enhancing code security is detailed in Table~\ref{tab:security}. 
Based on these results, we can make several key observations:
(1) \textbf{SFT significantly enhances code security.}
Across various LLMs, SFT training typically elevates baseline security rates from approximately 70.2\%-75.3\% to a substantially improved range of 84.2\%-90.3\%. 
In comparison, while DPO alone offers some improvements over the original models, SFT consistently provides more substantial security enhancements.
(2) \textbf{Subsequent DPO training after SFT offers limited additional benefits and can occasionally be detrimental.}
LLMs trained with SFT achieve security rates ranging from 84.2\% to 90.3\%. 
After subsequent DPO training, these rates adjust to a range of 84.7\% to 91.2\%. 
Notably, in some cases, the S\&D training leads to slight reductions in overall security rates.
For example, Qwen2.5-Coder 1.5B trained with SFT achieves a security rate of 88.9\%, which decreases to 88.3\% after subsequent DPO training.

\begin{table}[htbp]
  \centering
  \caption{RQ-1: The effectiveness of SFT and DPO in improving security rate}
  \resizebox{.7\linewidth}{!}
{
    \begin{tabular}{l|cccc}
    \toprule
    \textbf{LLM} & \textbf{Ori.} & \textbf{SFT} & \textbf{DPO} & \textbf{S\&D} \\
    \midrule
    Llama 1B & 77.4\% & 90.3\% & 80.9\% & \cellcolor[rgb]{ .886,  .937,  .855}\textbf{91.2\%} \\
    DeepSeek 1.3B & 71.7\% & 85.7\% & 73.3\% & \cellcolor[rgb]{ .886,  .937,  .855}\textbf{86.4\%} \\
    Qwen 1.5B & 75.3\% & \cellcolor[rgb]{ .886,  .937,  .855}\textbf{88.9\%} & 76.6\% & 88.3\% \\
    \midrule
    Magicoder 6.7B & 71.2\% & 84.2\% & 72.4\% & \cellcolor[rgb]{ .886,  .937,  .855}\textbf{85.1\%} \\
    DeepSeek 6.7B & 72.1\% & \cellcolor[rgb]{ .886,  .937,  .855}\textbf{84.9\%} & 72.7\% & 84.7\% \\
    Qwen 7B & 70.2\% & \cellcolor[rgb]{ .886,  .937,  .855}\textbf{86.8\%} & 71.9\% & 86.4\% \\
    \bottomrule
    \end{tabular}%
}
  \label{tab:security}%
\end{table}%

We further compare SFT and S\&D across different Common Weakness Enumerations (CWEs).
We focus on the five most frequent CWEs in the CyberSec~\cite{wan2024cyberseceval} benchmark: CWE-78, CWE-94, CWE-328, CWE-502, and CWE-798. 
As shown in Table~\ref{tab:cwe}, SFT and S\&D each demonstrate advantages on different CWEs, but the overall differences are not significant.

\begin{table}[htbp]
    \centering
    \caption{RQ-1: The performance comparison among six LLMs on Top-5 CWEs (Security / Total)}
    \label{tab:cwe}%
    \resizebox{.95\linewidth}{!}
    {
      \begin{tabular}{lcc|cc|cc}
      \toprule
      \multirow{1.5}[4]{*}{\textbf{CWE}} & \multicolumn{2}{c|}{\textbf{Llama 1B}} & \multicolumn{2}{c|}{\textbf{DeepSeek 1.3B}} & \multicolumn{2}{c}{\textbf{Qwen 1.5B}} \\
  \cmidrule{2-7}      & \textbf{SFT} & \textbf{S\&D} & \textbf{SFT} & \textbf{S\&D} & \textbf{SFT} & \textbf{S\&D} \\
      \midrule
      CWE-78 & \cellcolor[rgb]{ .996,  .78,  .812}373/415 & \cellcolor[rgb]{ .886,  .937,  .855}376/415 & \cellcolor[rgb]{ .996,  .78,  .812}367/415 & \cellcolor[rgb]{ .886,  .937,  .855}373/415 & \cellcolor[rgb]{ .886,  .937,  .855}388/415 & \cellcolor[rgb]{ .996,  .78,  .812}386/415 \\
      CWE-94 & \cellcolor[rgb]{ .886,  .937,  .855}300/325 & \cellcolor[rgb]{ .996,  .78,  .812}295/325 & \cellcolor[rgb]{ .996,  .78,  .812}244/325 & \cellcolor[rgb]{ .886,  .937,  .855}254/325 & \cellcolor[rgb]{ .886,  .937,  .855}279/325 & \cellcolor[rgb]{ .996,  .78,  .812}271/325 \\
      CWE-328 & \cellcolor[rgb]{ .996,  .78,  .812}256/275 & \cellcolor[rgb]{ .886,  .937,  .855}269/275 & \cellcolor[rgb]{ .996,  .78,  .812}262/275 & \cellcolor[rgb]{ .886,  .937,  .855}270/275 & \cellcolor[rgb]{ .886,  .937,  .855}265/275 & \cellcolor[rgb]{ .996,  .78,  .812}264/275 \\
      CWE-502 & \cellcolor[rgb]{ .996,  .78,  .812}190/215 & \cellcolor[rgb]{ .886,  .937,  .855}194/215 & \cellcolor[rgb]{ .996,  .78,  .812}182/215 & \cellcolor[rgb]{ .886,  .937,  .855}187/215 & \cellcolor[rgb]{ .886,  .937,  .855}181/215 & \cellcolor[rgb]{ .886,  .937,  .855}181/215 \\
      CWE-798 & \cellcolor[rgb]{ .996,  .78,  .812}162/185 & \cellcolor[rgb]{ .886,  .937,  .855}166/185 & \cellcolor[rgb]{ .886,  .937,  .855}173/185 & \cellcolor[rgb]{ .996,  .78,  .812}162/185 & \cellcolor[rgb]{ .886,  .937,  .855}168/185 & \cellcolor[rgb]{ .996,  .78,  .812}166/185 \\
      \midrule
      \multirow{1.5}[4]{*}{\textbf{CWE}} & \multicolumn{2}{c|}{\textbf{Magicoder 6.7B}} & \multicolumn{2}{c|}{\textbf{DeepSeek 6.7B}} & \multicolumn{2}{c}{\textbf{Qwen 7B}} \\
  \cmidrule{2-7}      & \textbf{SFT} & \textbf{S\&D} & \textbf{SFT} & \textbf{S\&D} & \textbf{SFT} & \textbf{S\&D} \\
      \midrule
      CWE-78 & \cellcolor[rgb]{ .886,  .937,  .855}383/415 & \cellcolor[rgb]{ .996,  .78,  .812}371/415 & \cellcolor[rgb]{ .996,  .78,  .812}371/415 & \cellcolor[rgb]{ .886,  .937,  .855}373/415 & \cellcolor[rgb]{ .886,  .937,  .855}380/415 & \cellcolor[rgb]{ .996,  .78,  .812}377/415 \\
      CWE-94 & \cellcolor[rgb]{ .996,  .78,  .812}250/325 & \cellcolor[rgb]{ .886,  .937,  .855}261/325 & \cellcolor[rgb]{ .886,  .937,  .855}266/325 & \cellcolor[rgb]{ .996,  .78,  .812}259/325 & \cellcolor[rgb]{ .886,  .937,  .855}280/325 & \cellcolor[rgb]{ .996,  .78,  .812}272/325 \\
      CWE-328 & \cellcolor[rgb]{ .996,  .78,  .812}235/275 & \cellcolor[rgb]{ .886,  .937,  .855}246/275 & \cellcolor[rgb]{ .996,  .78,  .812}233/275 & \cellcolor[rgb]{ .886,  .937,  .855}242/275 & \cellcolor[rgb]{ .996,  .78,  .812}254/275 & \cellcolor[rgb]{ .886,  .937,  .855}264/275 \\
      CWE-502 & \cellcolor[rgb]{ .886,  .937,  .855}182/215 & \cellcolor[rgb]{ .996,  .78,  .812}180/215 & \cellcolor[rgb]{ .996,  .78,  .812}180/215 & \cellcolor[rgb]{ .886,  .937,  .855}181/215 & \cellcolor[rgb]{ .996,  .78,  .812}174/215 & \cellcolor[rgb]{ .886,  .937,  .855}175/215 \\
      CWE-798 & \cellcolor[rgb]{ .996,  .78,  .812}156/185 & \cellcolor[rgb]{ .886,  .937,  .855}161/185 & \cellcolor[rgb]{ .886,  .937,  .855}168/185 & \cellcolor[rgb]{ .996,  .78,  .812}159/185 & \cellcolor[rgb]{ .886,  .937,  .855}153/185 & \cellcolor[rgb]{ .886,  .937,  .855}153/185 \\
      \bottomrule
      \end{tabular}%
    }
  \end{table}%

\underline{\textbf{Code Smell Preference.}}
The effectiveness of SFT and DPO in reducing code smell is reported in Table~\ref{tab:smell}. 
Based on these results, we observe the following key findings:
(1) \textbf{SFT consistently outperforms DPO in reducing code smell.}
For example, with the Llama 3.2 1B model, SFT training enhances the Clean@5 score from 1.2\% to 4.6\%, whereas DPO training only increases it to 2.8\%.
Regarding clean rate, SFT training boosts it from 80.0\% to 95.5\%, while DPO training only increases it to 81.8\%.
(2) \textbf{Applying DPO training after SFT often proves counterproductive.} 
LLMs trained with SFT achieve clean rates ranging from 91.2\% to 96.6\%; however, the S\&D training reduces this range to 88.4\% to 95.4\%.
(3) \textbf{Overall, SFT demonstrates substantial improvements across all models.}
SFT improves Clean@5 scores from an initial range of 1.2\%-18.0\% to a resulting range of 4.6\%-23.6\%, and enhances clean rates from 75.6\%-85.7\% to 90.1\%-96.6\%.

\begin{table}[htbp]
\centering
\caption{RQ-1: The effectiveness of SFT and DPO in reducing code smell}
\label{tab:smell}%
\resizebox{\linewidth}{!}
    {
    \begin{tabular}{l|rrrr|rrrr}
    \toprule
    \multirow{1.5}[4]{*}{\textbf{LLM}} & \multicolumn{4}{c|}{\textbf{Clean@5}} & \multicolumn{4}{c}{\textbf{Clean Rate}} \\
    \cmidrule{2-9}      & \textbf{Ori.} & \textbf{SFT} & \textbf{DPO} & \textbf{S\&D} & \textbf{Ori.} & \textbf{SFT} & \textbf{DPO} & \multicolumn{1}{c}{\textbf{S\&D}} \\
    \midrule
    Llama 1B & 1.2\% & \cellcolor[rgb]{ .886,  .937,  .855}\textbf{4.6\%} & 2.8\% & 3.6\% & 80.0\% & \cellcolor[rgb]{ .886,  .937,  .855}\textbf{95.5\%} & 81.8\% & 95.4\% \\
    DeepSeek 1.3B & 7.4\% & \cellcolor[rgb]{ .886,  .937,  .855}\textbf{8.2\%} & 6.0\% & 7.4\% & 75.6\% & \cellcolor[rgb]{ .886,  .937,  .855}\textbf{90.1\%} & 74.6\% & 89.3\% \\
    Qwen 1.5B & 8.0\% & \cellcolor[rgb]{ .886,  .937,  .855}\textbf{11.8\%} & 8.8\% & 10.8\% & 81.2\% & \cellcolor[rgb]{ .886,  .937,  .855}\textbf{96.6\%} & 82.8\% & 93.4\% \\
    \midrule
    Magicoder 6.7B & 15.0\% & \cellcolor[rgb]{ .886,  .937,  .855}\textbf{20.4\%} & 13.4\% & 19.8\% & 85.7\% & \cellcolor[rgb]{ .886,  .937,  .855}\textbf{92.4\%} & 80.7\% & 88.4\% \\
    DeepSeek 6.7B & 15.0\% & \cellcolor[rgb]{ .886,  .937,  .855}\textbf{18.2\%} & 14.8\% & 17.4\% & 83.8\% & 91.2\% & 83.6\% & \cellcolor[rgb]{ .886,  .937,  .855}\textbf{92.0\%} \\
    Qwen 7B & 18.0\% & \cellcolor[rgb]{ .886,  .937,  .855}\textbf{23.6\%} & 23.0\% & 22.6\% & 82.1\% & \cellcolor[rgb]{ .886,  .937,  .855}\textbf{94.8\%} & 84.9\% & 94.7\% \\
    \bottomrule
    \end{tabular}%
    }
\end{table}%

\intuition{
{\bf Answer to RQ-1:}
For scenario \ding{182}, SFT is sufficient for achieving strong performance, while S\&D may not provide additional benefits and may even lead to diminished results.
}
\label{sec:rq1}

\subsection{RQ-2: Comparable Study of SFT and DPO on Scenario \ding{183}}

\noindent
\textbf{Objective.}
In Section~\ref{sec:understand}, we hypothesize that for scenario \ding{183}, sequential application of SFT followed by DPO (S\&D) achieves optimal performance, where SFT rapidly builds fundamental capabilities and DPO subsequently enables exploration of superior solutions.
This research question aims to empirically evaluate SFT and DPO in improving code efficiency, code complexity, and code conciseness. 
Through this evaluation, we seek to validate our hypothesis and identify which strategy offers greater practical advantages.

\noindent
\textbf{Experimental Design.}
The models and training procedures used in RQ-2 are consistent with those in RQ-1 (refer to Section~\ref{sec:rq1} for detailed specifications). 
The evaluation metrics for code efficiency, code complexity, and code conciseness are \textit{Efficient@5}, \textit{Simple@5}, and \textit{Concise@5}, respectively, as defined in Section~\ref{sec:experiment}. 
Additionally, we calculate the \textit{Efficiency Rate}, \textit{Simplicity Rate}, and \textit{Conciseness Rate}, which represent the proportion of code that meets the corresponding preference standards among the solutions that pass all test cases.

\noindent
\textbf{Results.}
We present and analyze the results for code efficiency, complexity, and conciseness preferences, respectively.

\underline{\textbf{Code Efficiency Preference.}}
The impact of SFT, DPO, and the combined S\&D training on improving code efficiency is detailed in Table~\ref{tab:efficiency}. 
We observe that models first establish foundational capabilities in generating efficient code after SFT training. 
For instance, SFT substantially improves both Efficient@5 and efficiency rate metrics across all models compared to their original baselines (e.g., for Llama 1B, Efficient@5 improves from 0.8\% to 3.0\%, and efficiency rate increases from 50.0\% to 85.7\%).
Building upon this SFT-established foundation, the subsequent application of DPO training, as part of the S\&D training, appears to further encourage the models to explore and generate more efficient code. 
Consequently, the S\&D training frequently achieves superior performance, demonstrating advantages in both Efficient@5 and efficiency rate metrics across several LLMs.

\begin{table}[htbp]
\centering
\caption{RQ-2: The effectiveness of SFT and DPO in improving code efficiency}
\resizebox{\linewidth}{!}
{
    \begin{tabular}{l|rrrr|rrrr}
    \toprule
    \multirow{1.5}[4]{*}{\textbf{LLM}} & \multicolumn{4}{c|}{\textbf{Efficient@5}} & \multicolumn{4}{c}{\textbf{Efficiency Rate}} \\
    \cmidrule{2-9}      & \textbf{Ori.} & \textbf{SFT} & \textbf{DPO} & \textbf{S\&D} & \textbf{Ori.} & \textbf{SFT} & \textbf{DPO} & \multicolumn{1}{c}{\textbf{S\&D}} \\
    \midrule
    Llama 1B & 0.8\% & \cellcolor[rgb]{ .886,  .937,  .855}\textbf{3.0\%} & 0.8\% & 2.8\% & 50.0\% & 85.7\% & 57.1\% & \cellcolor[rgb]{ .886,  .937,  .855}\textbf{93.8\%} \\
    DeepSeek 1.3B & 6.4\% & 7.0\%  & 6.6\% & \cellcolor[rgb]{ .886,  .937,  .855}\textbf{7.4\%}& 56.4\% & 77.1\% & \cellcolor[rgb]{ .886,  .937,  .855}\textbf{85.9\%} & 80.7\% \\
    Qwen 1.5B & 7.8\% & 9.8\% & 9.6\% & \cellcolor[rgb]{ .886,  .937,  .855}\textbf{10.0\%} & 74.1\% & 85.9\% & 83.3\% & \cellcolor[rgb]{ .886,  .937,  .855}\textbf{90.4\%} \\
    \midrule
    Magicoder 6.7B & 14.0\% & \cellcolor[rgb]{ .886,  .937,  .855}\textbf{21.0\%} & 14.2\% & 18.4\% & 77.1\% & 83.7\% & 78.2\% & \cellcolor[rgb]{ .886,  .937,  .855}\textbf{85.3\%} \\
    DeepSeek 6.7B & 15.4\% & 17.6\% & 16.6\% & \cellcolor[rgb]{ .886,  .937,  .855}\textbf{17.8\%} & 80.9\% & \cellcolor[rgb]{ .886,  .937,  .855}\textbf{84.7\%} & 81.3\% & 81.0\% \\
    Qwen 7B & 18.4\% & 18.4\% & 17.8\% & \cellcolor[rgb]{ .886,  .937,  .855}\textbf{18.8\%} & 80.8\% & 84.5\% & 78.0\% & \cellcolor[rgb]{ .886,  .937,  .855}\textbf{85.9\%} \\
    \bottomrule
    \end{tabular}%
}
\label{tab:efficiency}%
\end{table}%

\underline{\textbf{Code Complexity Preference.}}
Table~\ref{tab:complexity} summarizes the effectiveness of SFT and DPO in reducing code complexity, as measured by cyclomatic complexity.
Both SFT and S\&D training substantially improve Simple@5 scores compared to baseline models.
Although SFT achieves superior performance on Simple@5 metrics, it does not demonstrate significant improvements in simplicity rate. 
We attribute this phenomenon to the influence of code correctness requirements, as Simple@5 requires solutions to be both simple and pass all test cases, while DPO may reduce the pass rate of generated code. 
Nevertheless, S\&D achieves higher simplicity rates, indicating that among solutions passing all test cases, those generated by S\&D generally exhibit lower cyclomatic complexity.

\begin{table}[htbp]
\centering
\caption{RQ-2: The effectiveness of SFT and DPO in reducing cyclomatic complexity}
\resizebox{\linewidth}{!}
{
    \begin{tabular}{l|rrrr|rrrr}
    \toprule
    \multirow{1.5}[4]{*}{\textbf{LLM}} & \multicolumn{4}{c|}{\textbf{Simple@5}} & \multicolumn{4}{c}{\textbf{Simplicity Rate}} \\
    \cmidrule{2-9}      & \textbf{Ori.} & \textbf{SFT} & \textbf{DPO} & \textbf{S\&D} & \textbf{Ori.} & \textbf{SFT} & \textbf{DPO} & \multicolumn{1}{c}{\textbf{S\&D}} \\
    \midrule
    Llama 1B & 0.2\% & \cellcolor[rgb]{ .886,  .937,  .855}\textbf{5.2\%} & 1.4\% & 4.2\% & 10.0\% & \cellcolor[rgb]{ .886,  .937,  .855}\textbf{100.0\%} & 35.0\% & \cellcolor[rgb]{ .886,  .937,  .855}\textbf{100.0\%} \\
    DeepSeek 1.3B & 3.2\% & \cellcolor[rgb]{ .886,  .937,  .855}\textbf{9.4\%} & 5.0\% & 8.0\% & 26.9\% & 96.3\% & 48.5\% & \cellcolor[rgb]{ .886,  .937,  .855}\textbf{97.1\%} \\
    Qwen 1.5B & 3.0\% & 11.6\% & 6.4\% & \cellcolor[rgb]{ .886,  .937,  .855}\textbf{12.4\%} & 28.2\% & \cellcolor[rgb]{ .886,  .937,  .855}\textbf{98.9\%} & 58.1\% & 98.3\% \\
    \midrule
    Magicoder 6.7B & 8.2\% & \cellcolor[rgb]{ .886,  .937,  .855}\textbf{21.8\%} & 12.4\% & 19.8\% & 45.7\% & 96.9\% & 68.9\% & \cellcolor[rgb]{ .886,  .937,  .855}\textbf{97.5\%} \\
    DeepSeek 6.7B & 7.4\% & \cellcolor[rgb]{ .886,  .937,  .855}\textbf{20.0\%} & 12.2\% & 19.8\% & 33.8\% & 98.7\% & 59.9\% & \cellcolor[rgb]{ .886,  .937,  .855}\textbf{98.0\%} \\
    Qwen 7B & 6.6\% & \cellcolor[rgb]{ .886,  .937,  .855}\textbf{22.8\%} & 18.6\% & 20.4\% & 30.0\% & 96.2\% & 63.2\% & \cellcolor[rgb]{ .886,  .937,  .855}\textbf{97.6\%} \\
    \bottomrule
    \end{tabular}%
}
\label{tab:complexity}%
\end{table}%

\underline{\textbf{Code Conciseness Preference.}}
Table~\ref{tab:conciseness} presents the results on code conciseness, evaluating metrics such as Concise@5 and conciseness rate after applying SFT and DPO training. 
The findings demonstrate that while SFT significantly enhances the conciseness of code generated by baseline LLMs, the S\&D training yields superior outcomes.
Notably, S\&D generally surpasses SFT in improving Concise@5 scores across several models. 
Furthermore, S\&D particularly excels in maximizing the conciseness rate, achieving top performance for all tested LLMs, with rates often approaching near-optimal levels (ranging from 93.0\% to 100.0\%). 
This underscores the effectiveness of the combined SFT and DPO training in guiding LLMs to produce highly concise code solutions.

\begin{table}[htbp]
\centering
\caption{RQ-2: The effectiveness of SFT and DPO in improving code conciseness}
\resizebox{\linewidth}{!}
{
    \begin{tabular}{l|rrrr|rrrr}
    \toprule
    \multirow{1.5}[4]{*}{\textbf{LLM}} & \multicolumn{4}{c|}{\textbf{Concise@5}} & \multicolumn{4}{c}{\textbf{Conciseness Rate}} \\
    \cmidrule{2-9}      & \textbf{Ori.} & \textbf{SFT} & \textbf{DPO} & \textbf{S\&D} & \textbf{Ori.} & \textbf{SFT} & \textbf{DPO} & \multicolumn{1}{c}{\textbf{S\&D}} \\
    \midrule
    Llama 1B & 0.4\% & \cellcolor[rgb]{ .886,  .937,  .855}\textbf{4.0\%} & 1.4\% & 3.8\% & 20.0\% & 96.7\% & 42.1\% & \cellcolor[rgb]{ .886,  .937,  .855}\textbf{100.0\%} \\
    DeepSeek 1.3B & 5.4\% & \cellcolor[rgb]{ .886,  .937,  .855}\textbf{6.6\%} & 4.2\% & 6.4\% & 51.3\% & 86.8\% & 51.8\% & \cellcolor[rgb]{ .886,  .937,  .855}\textbf{97.2\%} \\
    Qwen 1.5B & 5.0\% & 8.0\% & 5.8\% & \cellcolor[rgb]{ .886,  .937,  .855}\textbf{8.6\%} & 36.5\% & 92.1\% & 43.3\% & \cellcolor[rgb]{ .886,  .937,  .855}\textbf{93.0\%} \\
    \midrule
    Magicoder 6.7B & 12.0\% & 18.2\% & 13.0\% & \cellcolor[rgb]{ .886,  .937,  .855}\textbf{18.6\%} & 65.7\% & 93.4\% & 61.1\% & \cellcolor[rgb]{ .886,  .937,  .855}\textbf{95.0\%} \\
    DeepSeek 6.7B & 13.6\% & 15.8\% & 14.0\% & \cellcolor[rgb]{ .886,  .937,  .855}\textbf{16.4\%} & 66.7\% & 91.5\% & 64.6\% & \cellcolor[rgb]{ .886,  .937,  .855}\textbf{93.1\%} \\
    Qwen 7B & 11.4\% & 17.0\% & 16.2\% & \cellcolor[rgb]{ .886,  .937,  .855}\textbf{18.2\%} & 43.3\% & 94.9\% & 49.7\% & \cellcolor[rgb]{ .886,  .937,  .855}\textbf{97.7\%} \\
    \bottomrule
    \end{tabular}%
}
\label{tab:conciseness}%
\end{table}%

\begin{figure}
    \centering
    \includegraphics[width=.9\linewidth]{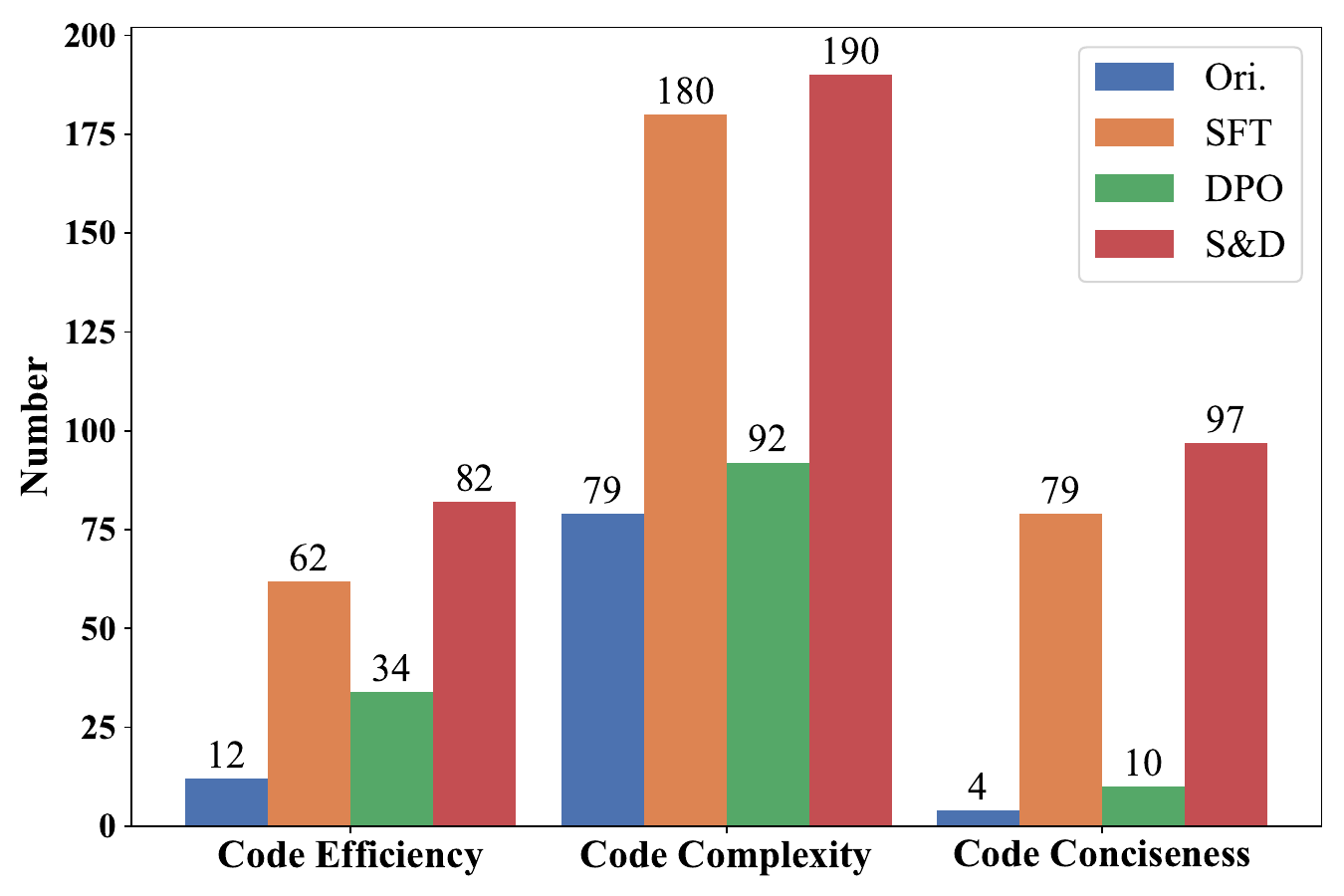}
    \caption{Number of best-performance instances across code preference scenarios}
    \label{fig:count}
\end{figure}

To further demonstrate the effectiveness of SFT and DPO training strategies, we conduct an additional comparative analysis. 
For each LLM, we first identify the common set of problems for which its original version, as well as its variants trained with SFT, DPO, and S\&D respectively, all generate solutions that successfully pass test cases. 
For each problem within this common set, we randomly select one solution from the pool generated by each of these four model variants (i.e., Original, SFT, DPO, and S\&D). 
After evaluating these selected solutions based on predefined performance metrics, we calculate the total number of times each distinct approach yields the best-performing solution.
As shown in Fig.~\ref{fig:count}, we observe that SFT establishes strong foundational capabilities. 
Subsequent DPO training (S\&D) then effectively promotes further exploration and refinement of solutions. 
Consequently, the S\&D training consistently achieves the highest number of best-performance instances, underscoring its overall superiority in these comparative evaluations.

\intuition{
{\bf Answer to RQ-2}: 
For scenario \ding{183}, sequential application of SFT followed by DPO (S\&D) achieves optimal performance, where SFT rapidly builds fundamental capabilities and DPO subsequently enables exploration of superior solutions.
}
\label{sec:rq2}

\subsection{RQ-3: Comparing APO with SFT and DPO}

\noindent
\textbf{Objective.}
While SFT effectively increases the probability assigned to preferred responses ($y^+$), DPO enables exploration of superior solutions and effectively reduces the probability of dispreferred responses ($y^-$), though it may also inadvertently decrease the probability of preferred responses ($y^+$).
Given that SFT and DPO demonstrate distinct advantages across different preference scenarios, we propose a unified approach that eliminates the need for manual scenario discrimination while simultaneously boosting the probability of preferred responses ($y^+$), suppressing dispreferred responses ($y^-$), and encouraging exploration of potentially superior solutions during training.
We term this dynamic integration framework \textbf{A}daptive \textbf{P}reference \textbf{O}ptimization (APO), which adaptively leverages the complementary strengths of SFT and DPO.
In this research question, we aim to determine whether our APO framework offers superior performance in optimizing response preferences and fostering solution exploration when compared to standalone SFT, DPO, and their sequential S\&D application.

\noindent
\textbf{Experimental Design.}
The models and metrics used in RQ-3 are consistent with those in RQ-1 and RQ-2. 
Considering the respective strengths of SFT and S\&D in different scenarios, we compare APO with SFT in scenario \ding{182} and with S\&D in scenario \ding{183}.
We also evaluate the efficiency of APO, SFT, DPO, and S\&D in terms of training time and GPU memory cost. 
For these analyses, we select three representative models: \texttt{Qwen2.5-Coder 7B}, \texttt{Magicoder 6.7B}, and \texttt{DeepSeek-Coder 6.7B}. 
Training time cost is measured as the average time required to complete all training epochs for each preference scenario, and GPU memory cost is reported as the average usage during training.

\noindent
\textbf{Results.} We present and analyze the results from both effectiveness and efficiency perspectives, respectively.

\begin{table}[htbp]
  \centering
  \caption{RQ-3: The effectiveness of APO and SFT in scenario \ding{182}}
  \resizebox{\linewidth}{!}
  {
    \begin{tabular}{l|cc|cc|cc|cc}
    \toprule
    \multirow{1.5}[4]{*}{\textbf{LLM}} & \multicolumn{2}{c|}{\textbf{Pass@5}} & \multicolumn{2}{c|}{\textbf{Security Rate}} & \multicolumn{2}{c|}{\textbf{Clean@5}} & \multicolumn{2}{c}{\textbf{Clean Rate}} \\
    \cmidrule{2-9}      & \textbf{SFT} & \textbf{APO} & \textbf{SFT} & \textbf{APO} & \textbf{SFT} & \textbf{APO} & \textbf{SFT} & \textbf{APO} \\
    \midrule
    Llama 1B & \cellcolor[rgb]{ .886,  .937,  .855}\textbf{6.2\%} & 3.8\% & 90.3\% & \cellcolor[rgb]{ .886,  .937,  .855}\textbf{90.9\%} & \cellcolor[rgb]{ .886,  .937,  .855}\textbf{4.6\%} & 4.4\% & 95.5\% & \cellcolor[rgb]{ .886,  .937,  .855}\textbf{100.0\%} \\
    DeepSeek 1.3B & \cellcolor[rgb]{ .886,  .937,  .855}\textbf{9.6\%} & \cellcolor[rgb]{ .886,  .937,  .855}\textbf{9.6\%} & \cellcolor[rgb]{ .886,  .937,  .855}\textbf{85.7\%} & 85.6\% & \textbf{8.2\%} & 6.0\% & 90.1\% & \cellcolor[rgb]{ .886,  .937,  .855}\textbf{95.2\%} \\
    Qwen 1.5B & \cellcolor[rgb]{ .886,  .937,  .855}\textbf{14.8\%} & 11.8\% & 88.9\% & \cellcolor[rgb]{ .886,  .937,  .855}\textbf{89.3\%} & \cellcolor[rgb]{ .886,  .937,  .855}\textbf{11.8\%} & 9.4\% & \cellcolor[rgb]{ .886,  .937,  .855}\textbf{96.6\%} & 94.7\% \\
    \midrule
    Magicoder 6.7B & \cellcolor[rgb]{ .886,  .937,  .855}\textbf{22.0\%} & \cellcolor[rgb]{ .886,  .937,  .855}\textbf{22.0\%} & 84.2\% & \cellcolor[rgb]{ .886,  .937,  .855}\textbf{84.9\%} & \cellcolor[rgb]{ .886,  .937,  .855}\textbf{20.4\%} & 17.0\% & 92.4\% & \cellcolor[rgb]{ .886,  .937,  .855}\textbf{97.5\%} \\
    DeepSeek 6.7B & \cellcolor[rgb]{ .886,  .937,  .855}\textbf{22.0\%} & 20.8\% & 84.9\% & \cellcolor[rgb]{ .886,  .937,  .855}\textbf{85.9\%} & \cellcolor[rgb]{ .886,  .937,  .855}\textbf{18.2\%} & 17.4\% & 91.2\% & \cellcolor[rgb]{ .886,  .937,  .855}\textbf{94.8\%} \\
    Qwen 7B & \cellcolor[rgb]{ .886,  .937,  .855}\textbf{27.0\%} & 24.8\% & \cellcolor[rgb]{ .886,  .937,  .855}\textbf{86.8\%} & 86.5\% & \cellcolor[rgb]{ .886,  .937,  .855}\textbf{23.6\%} & 20.0\% & 94.8\% & \cellcolor[rgb]{ .886,  .937,  .855}\textbf{96.4\%} \\
    \bottomrule
    \end{tabular}%
  }
  \label{tab:rq3_scenario_1}%
\end{table}%

\begin{table*}[htbp]
  \centering
  \caption{RQ-3: The effectiveness of APO and S\&D in scenario \ding{183}}
  \resizebox{.9\linewidth}{!}
  {
    \begin{tabular}{l|cc|cc|cc|cc|cc|cc}
    \toprule
    \multirow{1.5}[4]{*}{\textbf{LLM}} & \multicolumn{2}{c|}{\textbf{Efficient@5}} & \multicolumn{2}{c|}{\textbf{Efficiency Rate}} & \multicolumn{2}{c|}{\textbf{Simple@5}} & \multicolumn{2}{c|}{\textbf{Simplicity Rate}} & \multicolumn{2}{c|}{\textbf{Concise@5}} & \multicolumn{2}{c}{\textbf{Conciseness Rate}} \\
    \cmidrule{2-13}      & \textbf{S\&D} & \textbf{APO} & \textbf{S\&D} & \textbf{APO} & \textbf{S\&D} & \textbf{APO} & \textbf{S\&D} & \multicolumn{1}{c|}{\textbf{APO}} & \textbf{S\&D} & \textbf{APO} & \textbf{S\&D} & \textbf{APO} \\
    \midrule
    Llama 1B & 2.8\% & \cellcolor[rgb]{ .886,  .937,  .855}\textbf{3.1\%} & 93.8\% & \cellcolor[rgb]{ .886,  .937,  .855}\textbf{93.9\%} & 4.2\% & \cellcolor[rgb]{ .886,  .937,  .855}\textbf{5.2\%} & \cellcolor[rgb]{ .886,  .937,  .855}\textbf{100.0\%} & \cellcolor[rgb]{ .886,  .937,  .855}\textbf{100.0\%} & \cellcolor[rgb]{ .886,  .937,  .855}\textbf{3.8\%} & 3.2\% & \cellcolor[rgb]{ .886,  .937,  .855}\textbf{100.0\%} & 95.2\% \\
    DeepSeek 1.3B & \cellcolor[rgb]{ .886,  .937,  .855}\textbf{7.4\%} & 6.9\% & \cellcolor[rgb]{ .886,  .937,  .855}\textbf{80.7\%} & 76.0\% & 8.0\% & \cellcolor[rgb]{ .886,  .937,  .855}\textbf{9.0\%} & \cellcolor[rgb]{ .886,  .937,  .855}\textbf{97.1\%} & 96.8\% & 6.4\% & \cellcolor[rgb]{ .886,  .937,  .855}\textbf{7.2\%} & \cellcolor[rgb]{ .886,  .937,  .855}\textbf{97.2\%} & 96.1\% \\
    Qwen 1.5B & 10.0\% & \cellcolor[rgb]{ .886,  .937,  .855}\textbf{11.2\%} & 90.4\% & \cellcolor[rgb]{ .886,  .937,  .855}\textbf{91.7\%} & \cellcolor[rgb]{ .886,  .937,  .855}\textbf{12.4\%} & 12.2\% & 98.3\% & \cellcolor[rgb]{ .886,  .937,  .855}\textbf{99.3\%} & 8.6\% & \cellcolor[rgb]{ .886,  .937,  .855}\textbf{9.4\%} & 93.0\% & \cellcolor[rgb]{ .886,  .937,  .855}\textbf{98.0\%} \\
    \midrule
    Magicoder 6.7B & 18.4\% & \cellcolor[rgb]{ .886,  .937,  .855}\textbf{21.9\%} & \cellcolor[rgb]{ .886,  .937,  .855}\textbf{85.3\%} & 84.3\% & 19.8\% & \cellcolor[rgb]{ .886,  .937,  .855}\textbf{20.2\%} & 97.5\% & \cellcolor[rgb]{ .886,  .937,  .855}\textbf{98.2\%} & 18.6\% & \cellcolor[rgb]{ .886,  .937,  .855}\textbf{18.8\%} & 95.0\% & \cellcolor[rgb]{ .886,  .937,  .855}\textbf{96.9\%} \\
    DeepSeek 6.7B & 17.8\% & \cellcolor[rgb]{ .886,  .937,  .855}\textbf{18.0\%} & 81.0\% & \cellcolor[rgb]{ .886,  .937,  .855}\textbf{86.3\%} & \cellcolor[rgb]{ .886,  .937,  .855}\textbf{19.8\%} & 18.8\% & 98.0\% & \cellcolor[rgb]{ .886,  .937,  .855}\textbf{99.6\%} & 16.4\% & \cellcolor[rgb]{ .886,  .937,  .855}\textbf{17.8\%} & 93.1\% & \cellcolor[rgb]{ .886,  .937,  .855}\textbf{97.0\%} \\
    Qwen 7B & 18.8\% & \cellcolor[rgb]{ .886,  .937,  .855}\textbf{19.2\%} & 85.9\% & \cellcolor[rgb]{ .886,  .937,  .855}\textbf{90.8\%} & 20.4\% & \cellcolor[rgb]{ .886,  .937,  .855}\textbf{21.8\%} & 97.6\% & \cellcolor[rgb]{ .886,  .937,  .855}\textbf{99.4\%} & 18.2\% & \cellcolor[rgb]{ .886,  .937,  .855}\textbf{19.0\%} & \cellcolor[rgb]{ .886,  .937,  .855}\textbf{97.7\%} & 97.1\% \\
    \bottomrule
    \end{tabular}%
  }
  \label{tab:rq3_scenario_2}%
\end{table*}%

\underline{\textbf{Effectiveness Comparison.}}
The effectiveness of APO compared against SFT and S\&D is detailed in Tables~\ref{tab:rq3_scenario_1} and Table~\ref{tab:rq3_scenario_2}.
In scenario \ding{182}, as shown in Table~\ref{tab:rq3_scenario_1}, APO achieves performance outcomes that are broadly comparable to SFT. 
While SFT tends to yield higher scores in Pass@5 and Clean@5 metrics across several models, APO demonstrates competitive or superior results in security rate and clean rate metrics.

When evaluated against S\&D in scenario \ding{183} (Table~\ref{tab:rq3_scenario_2}), APO generally exhibits more pronounced advantages across multiple evaluation metrics. 
APO frequently achieves higher scores in Efficient@5 (e.g., 21.9\% for Magicoder 6.7B vs. S\&D's 18.4\%), Simple@5 (e.g., 21.8\% for Qwen 7B vs. S\&D's 20.4\%), and Concise@5 (e.g., 19.0\% for Qwen 7B vs. S\&D's 18.2\%).
These experimental results suggest that APO training effectively promotes model exploration towards qualitatively superior solutions across preference dimensions.

Overall, APO offers a significant advantage by streamlining the training pipeline for LLMs. 
Its unified approach can deliver comparable or superior capabilities without requiring practitioners to distinguish between specific use-case scenarios when choosing between SFT or S\&D, thereby simplifying both training and deployment processes.

\begin{table}[htbp]
  \centering
  \caption{RQ-3: The training time cost and average GPU memory cost on different tasks}
  \resizebox{\linewidth}{!}{
    \begin{tabular}{l|cccc}
      \toprule
      \textbf{Tasks: \textbf{Time Cost (Minutes)}} & \textbf{SFT} & \textbf{DPO} & \textbf{S\&D} & \textbf{APO} \\
      \midrule
      Code Correctness & 122 & 219 & 344 & 219 \\
      Code Security    & 122 & 209 & 331 & 209 \\
      Code Smell       & 116 & 555 & 895 & 555 \\
      Code Efficiency  & 134 & 256 & 338 & 260 \\
      Code Complexity  & 134 & 224 & 364 & 224 \\
      Code Conciseness & 166 & 236 & 398 & 231 \\
      \midrule
      \textbf{Average GPU Memory Cost (GB)} & 76 & 101 & 102 & 101 \\
      \bottomrule
    \end{tabular}
  }
  \label{tab:rq3_efficiency}
\end{table}

\underline{\textbf{Efficiency Comparison.}}
Table~\ref{tab:rq3_efficiency} presents detailed analyses of training time and GPU memory cost for different training strategies, including SFT, DPO, S\&D, and APO.
We observe that while APO generally requires more training time than SFT across the evaluated tasks, its training duration is notably similar to that of standalone DPO. 
More importantly, APO exhibits a clear time advantage over the S\&D pipeline, completing training significantly faster across all tasks (e.g., 219 minutes for APO vs. 344 minutes for S\&D on code correctness).
Regarding GPU memory cost, APO's average usage (101 GB) is comparable to DPO (101 GB) and S\&D (102 GB).
These efficiency characteristics position APO as a practical option for practical LLM training.

\intuition{
{\bf Answer to RQ-3}: 
APO provides a unified training framework that achieves comparable or superior performance to SFT and S\&D without requiring scenario-specific strategy selection, thereby simplifying the training pipeline.
Additionally, APO demonstrates competitive efficiency in terms of training time and GPU memory cost.
}
\label{sec:rq3}
\label{sec:results}

\section{Threats to Validity}

\noindent
\textbf{Internal Validity.}
Potential threats to internal validity primarily stem from the construction of our preference dataset and the design of evaluation metrics. 
Although we carefully filter and validate code solutions using rigorous test cases and established tools (e.g., \texttt{Pylint} and \texttt{Cirron}), there remains a risk of undetected errors or systematic biases in labeling code preferences. 
Additionally, our reliance on automated tools for code smell detection and security vulnerability injection may introduce measurement inaccuracies that could affect the validity of preference annotations.
To mitigate these risks, we employ multiple complementary models and evaluation metrics to triangulate results and reduce potential systematic biases.

\noindent
\textbf{External Validity.}
The generalizability of our findings may be constrained by several factors, including our choice of datasets, model architectures, and code preference scenarios. 
Our experiments are conducted primarily on the APPS dataset with a specific set of LLMs, which may not comprehensively represent the full spectrum of real-world programming tasks and model architectures encountered in practice.
To address these limitations, we include diverse model architectures across different parameter scales and evaluate multiple complementary preference criteria. 
However, future studies incorporating additional datasets, programming languages, and preference dimensions would be valuable to confirm the broader applicability and robustness of our paper.
\label{sec:threats}

\section{Related Work}

\subsection{Large Language Models for Code}
LLMs have shown strong capabilities in code generation due to their large-scale training on diverse datasets, such as CodeLlama~\cite{roziere2023code}, WizardCoder~\cite{luowizardcoder}, and DeepSeek-Coder~\cite{deepseek-coder}. 
These models are typically enhanced through instruction SFT to further optimize their code generation performance.
Given the challenge of collecting high-quality instructional data, researchers have increasingly adopted self-instruct methodologies to generate synthetic training data using powerful models like GPT-4~\cite{wang2022self,alpaca,codealpaca}.
Evol-Instruct~\cite{luowizardcoder} employs sophisticated prompting strategies to generate more complex and diverse training instances. 
OSS-Instruct~\cite{wei2024magicoder} enables LLMs to leverage real-world open-source code snippets, thereby improving the practical relevance and quality of generated solutions.
While SFT boosts code quality, its exclusive focus on correct examples limits its ability to teach preference discrimination, as models never encounter negative examples~\cite{hong2024orpo}.

\subsection{Preference Optimization for Code Models}
Recent researchers have employed Direct Preference Optimization (DPO)~\cite{rafailov2024direct} to align models using pairwise preference data. 
DPO enables models to learn ranking preferences and select superior solutions (e.g., more efficient code)~\cite{xu2024dpo,zhang2024plum,zhang2024codedpo}.
Several studies have explored preference optimization specifically for code generation.
Code-Optimize~\cite{gee2024code} constructs its training dataset from the MBPP-train subset, which comprises a limited set of 384 programming problems.
PLUM~\cite{zhang2024plum} leverages GPT-4 to generate comprehensive test cases for validating and ranking code solutions, currently achieving state-of-the-art performance in preference optimization for code models.
CodeDPO~\cite{zhang2024codedpo} uses a self-generation and validation mechanism to create balanced preference pairs, aiming to optimize both correctness and efficiency.

\label{sec:related_work}

\section{Conclusion and Future Work}
This paper systematically investigates the roles of SFT and DPO in aligning LLMs with diverse code preferences. 
Through theoretical analysis and empirical evidence, we demonstrate that SFT excels in scenarios with objectively verifiable optimal solutions, while S\&D enables superior exploration in scenarios without objectively verifiable optimal solutions. 
Based on these insights, we propose \textbf{A}daptive \textbf{P}reference \textbf{O}ptimization (APO), a unified framework that dynamically integrates SFT and DPO strengths without requiring manual strategy selection.
Extensive experiments across six representative code preference tasks validate our hypotheses and show that APO consistently matches or surpasses existing approaches while simplifying the training pipeline.
Our work provides both theoretical foundations and practical guidance for code preference alignment.
Future work will explore multi-turn programming scenarios and real-time human-in-the-loop alignment settings.
\label{sec:conclusion}

\balance
\bibliographystyle{IEEEtran}
\bibliography{main}

\end{document}